\documentclass[useAMS,usenatbib]{mn2e}

\newcommand{\msun}{M$_{\sun}$}

\newcommand{\msuns}{M$_{\sun}~$}

\newcommand{\nbody}{{\it n}-body$~$}

\newcommand{\gcm}{g~cm$^{-2}$}
\newcommand{\gcms}{g~cm$^{-2}$~}
\newcommand{\mj}{M$_{J}$}
\newcommand{\rj}{R$_{J}$}

\newcommand{\mnras}{MNRAS}
\newcommand{\icarus}{Icarus}
\newcommand{\apj}{ApJ}
\newcommand{\aaps}{A\&AS}
\newcommand{\apjl}{ApJ}
\newcommand{\aj}{AJ}
\newcommand{\aap}{A\&A}

\newcommand{\nat}{Nature}

\newcommand{\fargo}{{\sc fargo }}
\newcommand{\mercury}{{\sc mercury }}
\newcommand{\fargons}{{\sc fargo}}
\newcommand{\mercuryns}{{\sc mercury}}

\usepackage{graphicx}
\usepackage{epstopdf}
\usepackage{amssymb,amsmath}
\usepackage{subfigure}

\title[Hydrodynamics of planet scattering]{Hydrodynamic outcomes of planet scattering in transitional discs}
\author[N. Moeckel and P. J. Armitage]{Nickolas Moeckel$^{1}$\thanks{E-mail:
moeckel@ast.cam.ac.uk} and Philip J. Armitage$^{2,3}$\\
$^{1}$Institute of Astronomy, University of Cambridge, Madingley Road, Cambridge, CB3 0HA\\
$^{2}$JILA, 440 UCB, University of Colorado, Boulder, CO 80309-0440, USA\\
$^{3}$Department of Astrophysical and Planetary Sciences, University of Colorado, Boulder, USA
}
\begin{document}

\date{Accepted XXX. Received YYY; in original form ZZZ}

\pagerange{\pageref{firstpage}--\pageref{lastpage}} \pubyear{2011}

\maketitle

\label{firstpage}

\begin{abstract}
A significant fraction of unstable multiple planet systems are likely to scatter 
during the transitional disc phase as gas damping becomes ineffectual. 
Using a large ensemble of \fargo hydrodynamic simulations and \mercury \nbody integrations, 
we directly follow the dynamics of planet-disc and planet-planet interactions through 
the clearing phase and on through 50 Myr of planetary system evolution. Disc 
clearing is assumed to occur as a result of X-ray driven photoevaporation. We find 
that the hydrodynamic evolution of individual scattering systems is complex, and can 
involve phases in which massive planets orbit within eccentric gaps, or accrete 
directly from the disc without a gap. Comparing the results to a reference gas-free 
model, we find that the \nbody dynamics and hydrodynamics of scattering into one- and 
two-planet final states are almost identical. The eccentricity distributions in these 
channels are almost unaltered by the presence of gas. The hydrodynamic 
simulations, however, also form a population of low eccentricity three-planet systems in 
long-term stable configurations, which are not found in \nbody runs. The 
admixture of these systems results in modestly lower eccentricities in 
hydrodynamic as opposed to gas-free simulations.  The precise incidence of these three-planet systems is likely 
a function of the initial conditions; different planet setups (number or spacing) may change the quantitative character of 
this result.
We analyze the properties of surviving 
multiple planet systems, and show that only a small fraction (a few percent) 
enter mean-motion resonances after scattering, while a larger fraction 
form stable resonant chains and avoid scattering entirely. Our results remain  
consistent with the hypothesis that exoplanet eccentricity results from 
scattering, though the detailed agreement between observations 
and gas-free simulation results is likely coincidental. We discuss the prospects for 
further tests of scattering 
models by observing planets or non-axisymmetric gas structure in transitional 
discs.
\end{abstract}


\section{Introduction}

Prior to gap opening, gravitational interactions between planets and their surrounding gas discs act 
to efficienctly damp planetary eccentricity. This damping implies that the formation of 
giant planets via core accretion leads to initially circular orbits, and that post-formation 
dynamical effects must be sought to explain the observed eccentricity distribution of massive 
extrasolar planets \citep{butler06}. Dynamical instability in multi-planet systems, leading to orbit crossing
and the ejection or merger of some of the planets, is one way to achieve this \citep{rasio96,weidenschilling96,lin97}.
Extensive \nbody experiments have shown that with realistic mass functions, the dynamics of unstable two-planet 
\citep{ford08}, three-planet \citep{chatterjee08}, or richer systems \citep{juric08} can successfully reproduce 
the observed eccentricity distribution.

The good agreement between \nbody dynamical experiments and observations is somewhat 
surprising, as it is hard to envisage a consistent scattering scenario in which gas disc 
interactions are not important. The simplest model is one in which the typical planetary 
system forms with just two massive planets. In this case, the only unstable systems have  
planets that are separated by less than the threshold for Hill stability \citep{marchal82,gladman93}, 
and most of the unstable subset scatter on a time scale that is much shorter than the gas 
disc dispersal time \citep[$\sim 10^5$ {\rm yr}][]{wolk96}. 
Scattering is therefore likely to occur as soon as gas disc damping 
declines below some threshold level, and will be synchronized with the epoch of disc dispersal. 
Related arguments apply, less robustly, to richer planetary systems. For more than two planets there is 
no analytic criterion for stability, and relatively loosely packed initial configurations 
can delay the onset of scattering for tens or hundreds of Myr \citep{chatterjee08}. Absent 
fine tuning, however, it is not possible to construct initial conditions in which {\em all} 
of the instabilities occur on $> {\rm Myr}$ time scales, without over-producing stable low 
eccentricity systems that are not seen observationally. As the gas 
dissipates there should therefore be a significant number of systems primed to go unstable, scattering
in the presence of at least some remnant gas.

The large computational cost of hydrodynamics relative to gravity has necessarily motivated approximate  approaches that
marry disc physics to \nbody simulations of scattering. One method is to include an analytic  description of planet-disc
interactions as a force within an \nbody calculation. This has been used to study  the migration of planets into resonance
\citep[e.g.][]{lee02,snellgrove01}, and through resonance into scattering \citep{adams03,moorhead05,lee09}. Scattering with
more than two planets in an embedding disc has been studied by \citet{chatterjee08} and \citet{matsumura10}, with the disc
modeled in 1D and torques on the planets calculated from that density profile. There have also been preliminary studies that
include a full hydrodynamic treatment of the gas disc during active periods of planet scattering. \citet{moeckel08} performed
3D simulations starting from very unstable equal-mass, two-planet systems with a remnant gas disc exterior
to the planets' orbits. \citet{marzari10} performed 2D simulations studying migration and resonant trapping in
three-planet systems, some of which led naturally to scattering when resonance was broken. While both of these studies hinted
at possibly significant dynamical effects mediated by the gas, they were too limited in scope to attempt 
statistical comparison to gas-free results.

Our goal in this work is to directly quantify the impact of residual gas discs on the outcome of 
planet scattering. We assume that the interesting dynamics is coincident with the epoch of disc 
dispersal, and follow the planet-planet and planet-disc interactions through this transition 
phase using a 2D hydrodynamic code. We consider three-planet systems, setup almost identically 
to previous gas-free calculations \citep{chatterjee08,raymond10}, 
and model disc dispersal using a physically motivated photoevaporation model \citep{owen11}. The 
hydrodynamics are run long enough to take the systems through the transition phase into the 
pure \nbody regime, after which we switch to a standard integrator for a further 50 Myr of 
gravitational interactions.

\section{The simulations}
\label{simdetails}
\subsection{Initial conditions}
To facilitate direct comparison with previous \nbody results, we 
consider three-planet systems around stars with mass $M_{\star} = 1$ \msun, with the mass $m$ of each planet drawn randomly from a power law distribution $f(m) \propto m^{-1.1}$ in the range 0.3--5 \mj. The innermost planet is placed with a semi-major axis $a_1= 3$ au, and the spacing of the other planets is determined following the approach of \citet{chambers96}, with each adjacent planetary pair separated by a fixed number of mutual Hill radii $R_{Hij}$. That is, with 
\begin{equation}
R_{Hij} = \left( \frac{m_i + m_j}{3 M_{\star}} \right)^{1/3} \frac{a_i + a_j}{2},
\label{eq:Hill}
\end{equation}
the middle planet is placed at $a_2 = a_1 + K R_{H12}$, and the outer planet at $a_3 = a_2 + K R_{H23}$. The constant $K$ largely determines the timescale on which gravitational instability occurs \citep{chambers96,marzari02,chatterjee08}. We set $K$ at 
approximately the maximum value that does not overproduce stable systems, whose low final eccentricity would be inconsistent 
with exoplanet observations.
With our mass spectrum, experimentation showed that $K=4.0$ yielded a distribution of instability times with a median $t_i \sim 4\times 10^4$ yr, and $\sim 10$\% of the setups unscattered after 50 Myr\footnote{This instability timescale is about an order of magnitude larger than the result for $K=4.0$ with equal mass planets found by \citet{chatterjee08}, where for a given $K$ each planetary pair is at a fixed separation relative to the major mean motion resonances. With our mass spectrum and $K=4$, each adjacent pair is somewhere between roughly the 3:2 and 2:1 mean motion resonances.}. 

\begin{figure}
 \includegraphics[width=84mm]{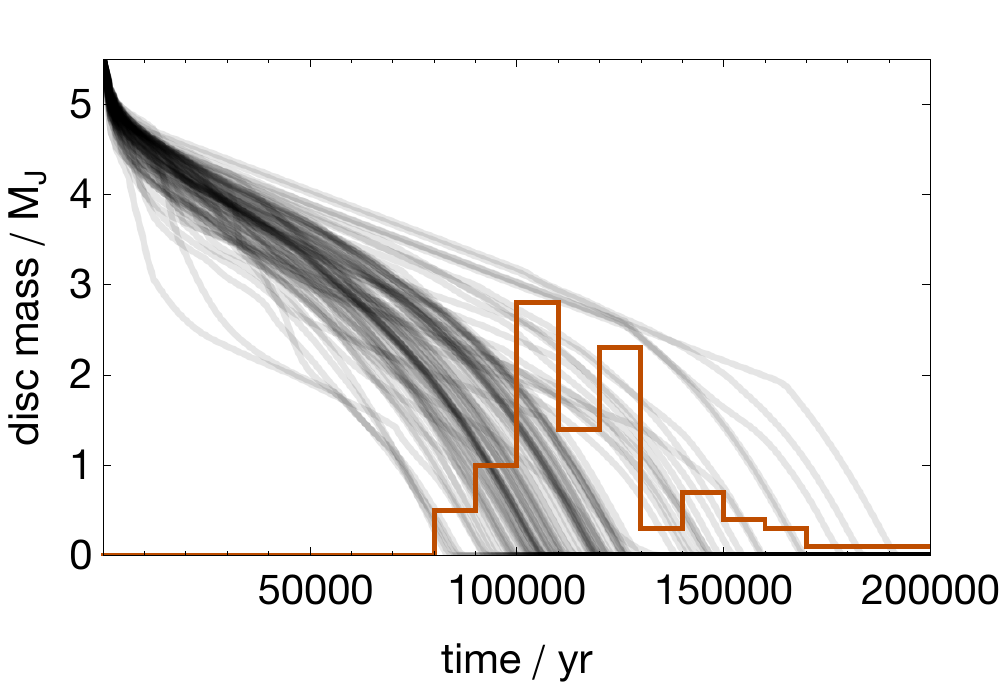}
 \caption{The evolution of the remaining disc mass for each run is plotted in gray. The red histogram shows the (arbitrarily normalized) distribution of dissipation times, i.e. when the disc mass reaches 0.}
 \label{dissipation}
\end{figure}

The planets are initially embedded in a gas disc which we model between 1 and 40 au. The surface density initially follows a radial profile $\Sigma(r) = 200 \ (r / 1 \ {\rm au})^{-1} \ {\rm g \ cm}^{-2}$. We include photoevaporation of the disc using the mass-loss profiles of \citet{owen11}, which are based on the coupled hydrodynamic and radiative transfer models introduced by \citet{owen10}. In these models, X-ray luminosity drives photoevaporative disc dispersal. Our discs initially contain just over 5 \mj~of gas, which is both the typical total mass of the three planets and the mass at which the photoevaporative mass loss begins to alter the disc structure on a fast timescale. 
We pragmatically set the X-ray luminosity of our star at $1.4\times10^{30}$ erg s$^{-1}$, which yields a disc clearing timescale of just over $10^5$ years depending on the details of the planet--disc interactions (see Figure \ref{dissipation}). This timescale, coupled with our chosen planet spacing, means that about half our systems might be expected to scatter when there is still a significant amount of disc material left.

We evolve 100 random realizations of the planetary initial conditions within the fixed initial conditions for the disc. We use direct hydrodynamic simulations to model the coupled planetary and disc dynamics while the disc mass is high enough to still influence the planets' evolution, followed by a longer \nbody run once the disc has been cleared. For the hydrodynamic portion of each run we use a modified version of the 2D code \fargo \citep{masset00}, and for the \nbody follow-on we use \mercury \citep{chambers99}.

\fargo is a 2D Eulerian hydrodynamic code tailored to planet-disc interactions \citep{masset00}. We have modified the code somewhat to make it more suitable for this work \citep[similar modifications were made by][]{marzari10}. The \nbody integrator was changed from a 5th order Runge-Kutta to a (7,8) order embedded pair \citep{prince81}. We include stepsize control in the following way: the results of the 7th and 8th order estimate of each component of each planet's velocity and position are compared at the end of each hydrodynamic timestep. If the fractional difference between the two exceeds $10^{-7}$, the planets are integrated through the hydrodynamic step with a smaller \nbody timestep, repeating this process until our accuracy criterion is met. 

We have also implemented the removal of escaping planets if they are on a hyperbolic orbit and more than 400 au from the star, as well as collisions between planets and between planets and the star. Planet--planet mergers are mass and momentum conserving and take place whenever two planets' surfaces are detected to touch, with all planets assumed to have radii of 1.3 \rj. Planets merge with the star if  they approach within 5 R$_{\sun}$.
\subsection{Numerical details}
\begin{figure}
 \includegraphics[width=84mm]{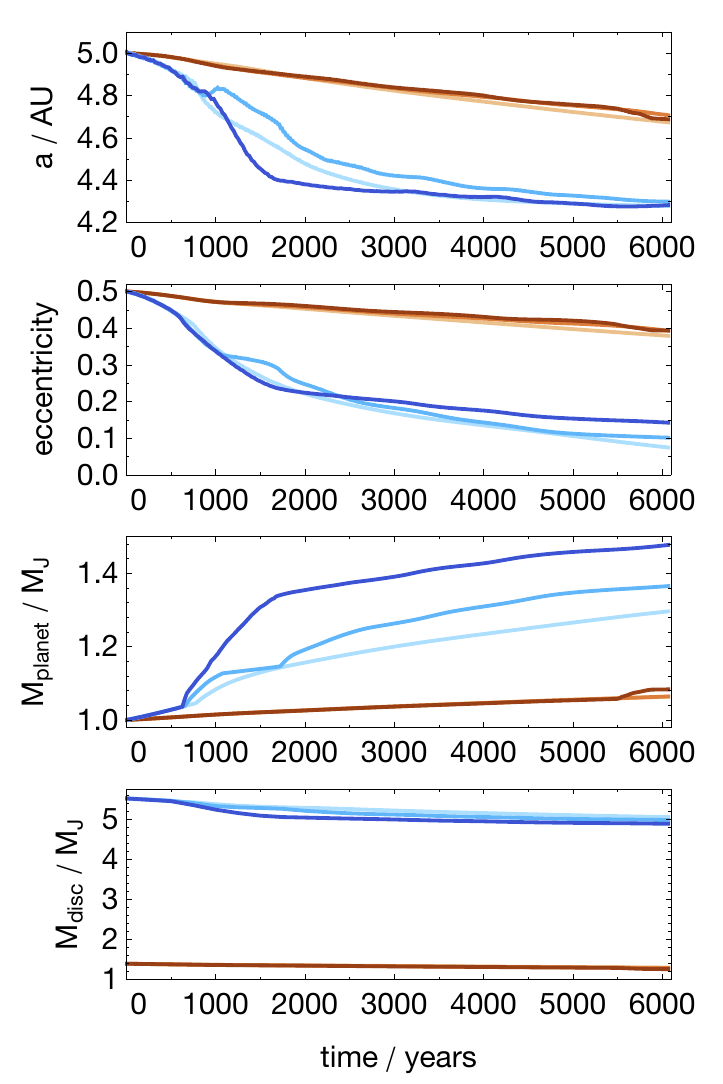}
 \caption{Resolution tests in which a 1 \mj~planet is placed on an eccentric orbit in an initially unperturbed disc. The series of blue lines shows a disc with $\Sigma(1~{\rm au}) = 200$ \gcm, and the red lines have $\Sigma(1~{\rm au}) = 50$ \gcm. In each series the lightest shade is run with $n_{\theta} = 256$, the medium shade with $n_{\theta}=512$, and the dark shade with $n_{\theta}=1024$. }
 \label{convergence}
\end{figure}

\begin{figure*}
 \includegraphics[width=180mm]{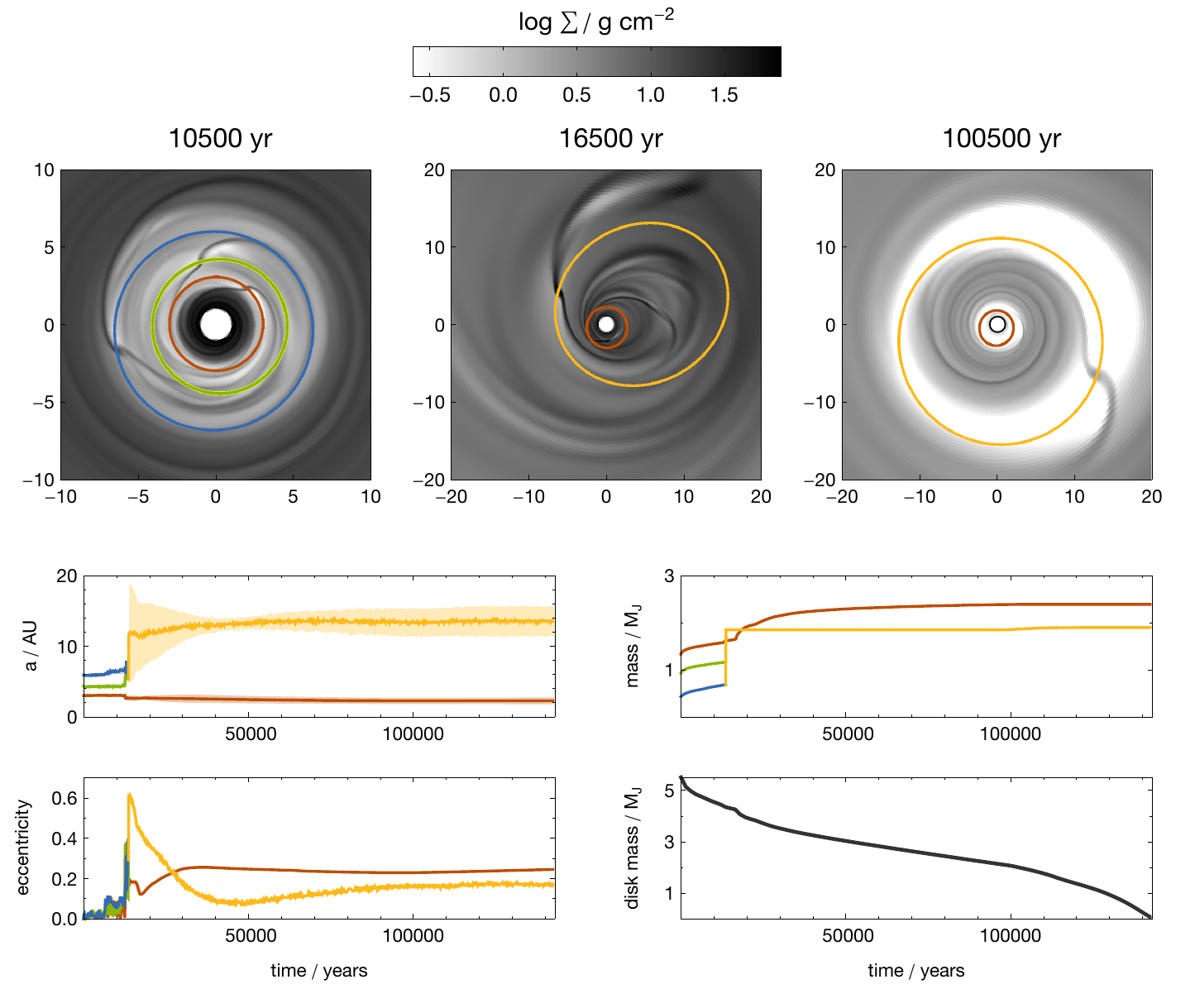}
 \caption{The evolution of the planetary system and the gas disc for an example run. The orbits of the planets are plotted on top of the surface density plots in the same colours as the plots of the orbital elements and masses. The surface density plot on the left shows the system before scattering, when the red, green, and blue planets have partially cleared a mutual gap. During the scattering event the green and blue planets collide to form the orange planet on an orbit with $e \sim 0.6$. The aftermath of the scattering is shown in the middle density plot.  Over the next $\sim 2.5\times 10^4$ years, disc interactions damp the eccentricity of the outer planet; after disc clearing, the two planets are stable. The right density plot shows the system near the end of the photoevaporative clearing, with each planet orbiting in a gap. }
  \label{example1}
\end{figure*}

\begin{figure*}
 \includegraphics[width=180mm]{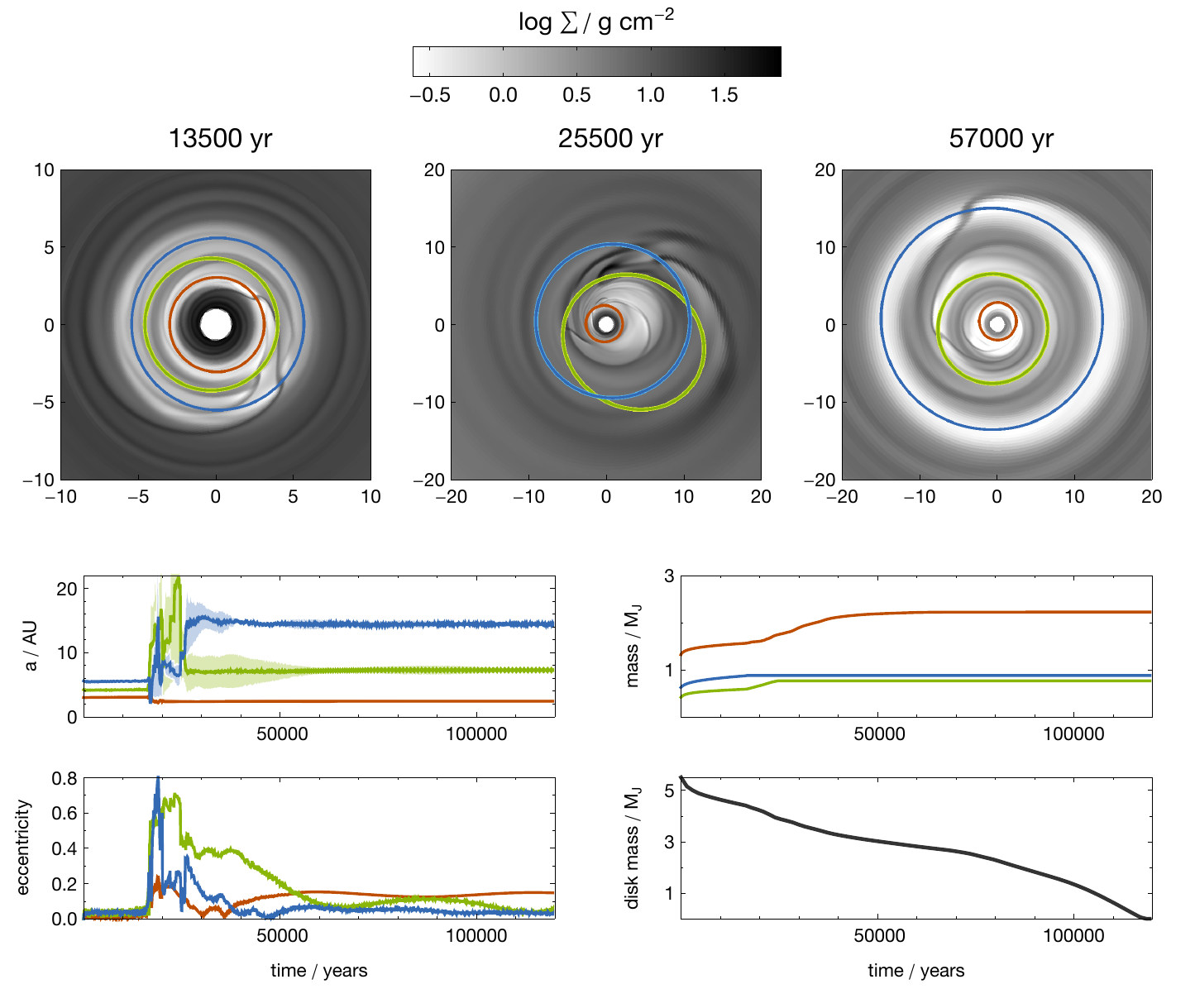}
 \caption{The evolution of the planetary system and the gas disc for an example run. The orbits of the planets are plotted on top of the surface density plots in the same colours as the plots of the orbital elements and masses. The surface density plot on the left shows the system before scattering, when the red, green, and blue planets have partially cleared a mutual gap. The middle density plot is during a period of active planetary dynamics.  After this stage the disc damps the eccentricity of the three planets, which remain in roughly this configuration through $10^8$ yr. This phase is shown in the right density plot.}
  \label{example2}
\end{figure*}
Photoevaporation is included using the X-ray driven mass-loss profile of \citet{owen11}, scaled to our 1 \msuns star and chosen X-ray luminosity of $1.4\times10^{30}$ erg s$^{-1}$. The azimuthally constant but radially dependent mass loss rate from these profiles is imposed across the numerical grid as an externally specified sink term in the continuity equation. In these models photoevaporation opens a gap near 2 au, and when the gas interior to the gap is gone the directly-illuminated outer disc erodes from the inside out \citep[see Figure 9 of][]{owen11}. We implement this transition to direct illumination somewhat approximately, as we do not follow the disc inwards of 1 au and set a lower surface density limit of $2\times10^{-3}$ \gcms below which we do not photoevaporate any material for numerical reasons\footnote{With this density floor, our final disc mass when photoevaporation has run its course is $\sim 10^{-3}$ \mj.}. As the simulation progresses, we track the azimuthally averaged surface density of each ring. When the innermost radius has an average surface density below $8\times10^{-3}$ \gcm, we rescale the inner radius of the photoevaporation profile to that radius. 

We allow accretion onto the planets using a similar approach to the unmodified version of \fargons, based on the prescription used by \citet{kley99}. The planet's Roche lobe (approximated by a constant radius $R_r$) is drained on some timescale, with material inside 0.45 $R_r$ drained at twice the rate of that between 0.45 and 0.75 $R_r$. Material outside 0.75 $R_r$ is left alone. The timescale on which the Roche lobe is drained is set to twice the planet's orbital period. With our chosen parameters, the disc is typically dispersed via a combination of planetary dynamics and photoevaporation (with a small contribution from accretion) between 1--1.5$\times 10^5$ yr, with a few instances stretching to 0.8--1.9$\times 10^5$ yr. In Figure \ref{dissipation} we show the mass evolution of all the discs in our runs.

We use open boundaries at the inner and outer edges of the grid. We use a fixed disc aspect ratio of $H/r = 0.04$, and an alpha-viscosity with $\alpha = 10^{-2}$. The grid is setup with $n_{\theta} = 512$ azimuthal cells, and $n_r =207$ rings. The radial spacing is logarithmic, with approximately square cells between 2.5 and 10 au, but coarser radial spacing between 1--2.5 and 10--40 au so that those cells are rectangular with a 2:1 aspect ratio. Our highest resolution is thus concentrated in the region of highest interest, where the majority of the planets spend most of the time. 

We tested the adequacy of the grid resolution by setting up a planet with $m=1$ \mj, and eccentricity $e=0.5$, in an initially unperturbed disc. We considered two initial disc masses, $\Sigma(1~{\rm au}) = 200$ and 50 \gcm, and three resolutions, $n_{\theta} = 256$, 512, and 1024. The results are shown in Figure \ref{convergence}. The more massive disc results are shown in blue, and the low mass disc in red. The lightest shade is $n_{\theta} = 256$, and the darkest shade is $n_{\theta}=1024$. The low mass disc is well converged in both the orbital evolution and planet mass. The high mass case is only moderately converged. While the orbit of the planet ends up at approximately the same place for all three resolutions, the mass evolution of the planet is quite different. The differences in the orbital evolution are most pronounced at the time when the planet is accreting the most mass, between 1000--2000 yr. Visual inspection of the surface density of these runs shows that gas is accreting onto the planet in streams that are not well resolved at low $n_{\theta}$. 
We conclude that for studying the orbital evolution of the planets --- the prime focus of this work --- $n_{\theta}=512$ is sufficient. Higher resolution, along with a more physical treatment of mass accretion within the Roche lobe, would likely be needed to model mass evolution in detail.

The results of the resolution test suggest that a massive planet scattering into an undisturbed disc will accrete material at a resolution-dependent rate until a gap is opened. This manifests itself as an offset in the mass evolution of the disc and the planet (see the similar slopes of the $M_{planet}$ and $M_{disc}$ curves in Figure \ref{convergence} after $\sim 1500$ years). Runs that scatter very early, when a large amount of gas is still in the disc, are susceptible to this numerical effect, which may alter the disc dissipation timescale. Offsets of the order 0.1 \mj~will not move the disc mass evolution out of the main band seen in Figure \ref{dissipation} though, and are certainly small compared
to the expected physical dispersion in dissipation history that results from varying stellar X-ray luminosity. Hence, we do not consider this to be of great importance.

Each set of initial conditions is integrated with \fargo until the disc is photoevaporated down to our density floor, at which point the run is terminated. The planetary system is then evolved further using the \fargo integrator with no disc until $2\times 10^5$ yr. At this point, the system is turned over to the Burlisch-Stoer integrator of \mercury \citep{chambers99} and integrated to 50 Myr. As in the \fargo portion of the integration, planetary mergers and ejections are allowed with the same parameters. 
These hydrodynamic runs are then compared to an ensemble of 500 purely \nbody simulations, set up using 
identical initial conditions but integrated using \mercury throughout.

\section{Results}
\label{results}
The hydrodynamic runs are characterised by several phases: an initial clearing of gaps and relaxation of the disc, the onset (or not) of dynamical instability and planet scattering, renewed relaxation of the disc and eccentricity damping, and clearing of the disc by photoevaporation. These phases are illustrated for two example runs in Figures \ref{example1} and \ref{example2}. In each of these examples scattering occurs, and the once-prominent gaps are filled in. As the disc comes to terms with the new planetary configuration eccentric gaps develop \citep[also noted by][]{marzari10}, and finally the disc is cleared leading to the purely gravitational phase of our experiments. Here we discuss the most interesting aspects of these simulations.

\subsection{Instability timescale and outcome}
\label{instabilitytimesection}
\begin{figure}
 \includegraphics[width=84mm]{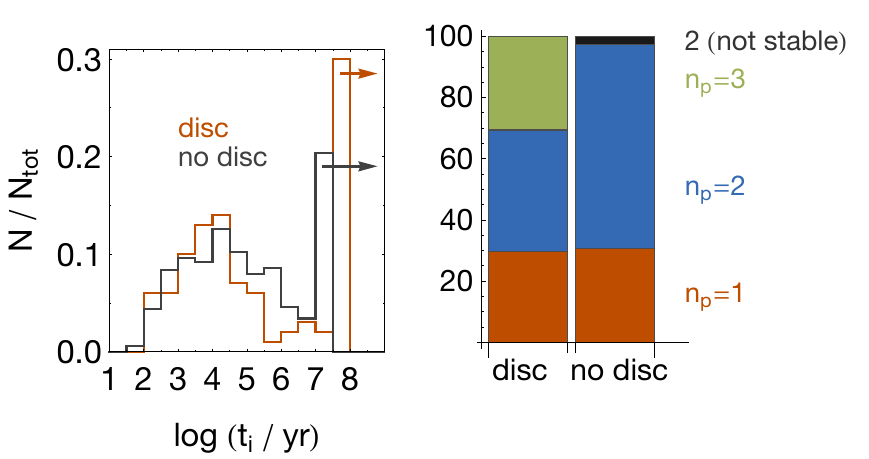}
 \caption{{\it Left:} The distribution of the instability times for the runs with a disc (red) and without (black). The final bins include all runs still stable at 50 Myr for the disc runs, and 15 Myr for the discless runs. {\it Right}:The final state of the planetary systems after 50 Myr (for the disc runs) and 15 Myr (for the discless runs). The bins labeled `$n_p = 1$' and  `$n_p = 2$' refer to single planet systems and double systems that are formally Hill stable. The bin labeled `2 (not stable)' refers to 2 planet systems that are not Hill stable. Triple planet systems (`$n_p=3$') include only those that underwent a scattering event but retain three planets at the end of the simulation.}
 \label{instability_times}
\end{figure}

In Figure \ref{instability_times} we compare the distribution of instability times between the hydrodynamic and \nbody control runs. During the \fargo runs, the time of scattering is identified as the first time the semi-major axis of any planet changes by 10 percent between two output times. Visual inspection of the results confirms that this is a reasonable estimate of the onset of instability. During the \mercury runs, an encounter between two planets within 3 Hill radii is the criterion. The bulk of the distribution is similar for both sets of simulations; the main difference is at the long-time tail of the distribution, where the \fargo runs have roughly 50\% more runs that never go unstable over the course of the simulations. 

In the right panel of the figure we plot the percentage of unstable runs ending in a given planetary configuration. The main difference between the simulation sets is the number of unscattered systems and the presence, in the hydrodynamic runs only, of a population of scattered systems that retain all three planets. There are also a handful of double systems that are not provably Hill stable in the \mercury runs, but not enough to alter any conclusions. We discuss the difference in the percentage of unscattered triple systems further in sections \ref{MMR2} and \ref{MMR3}.

\subsection{Eccentricity distributions}
\begin{figure}
 \includegraphics[width=84mm]{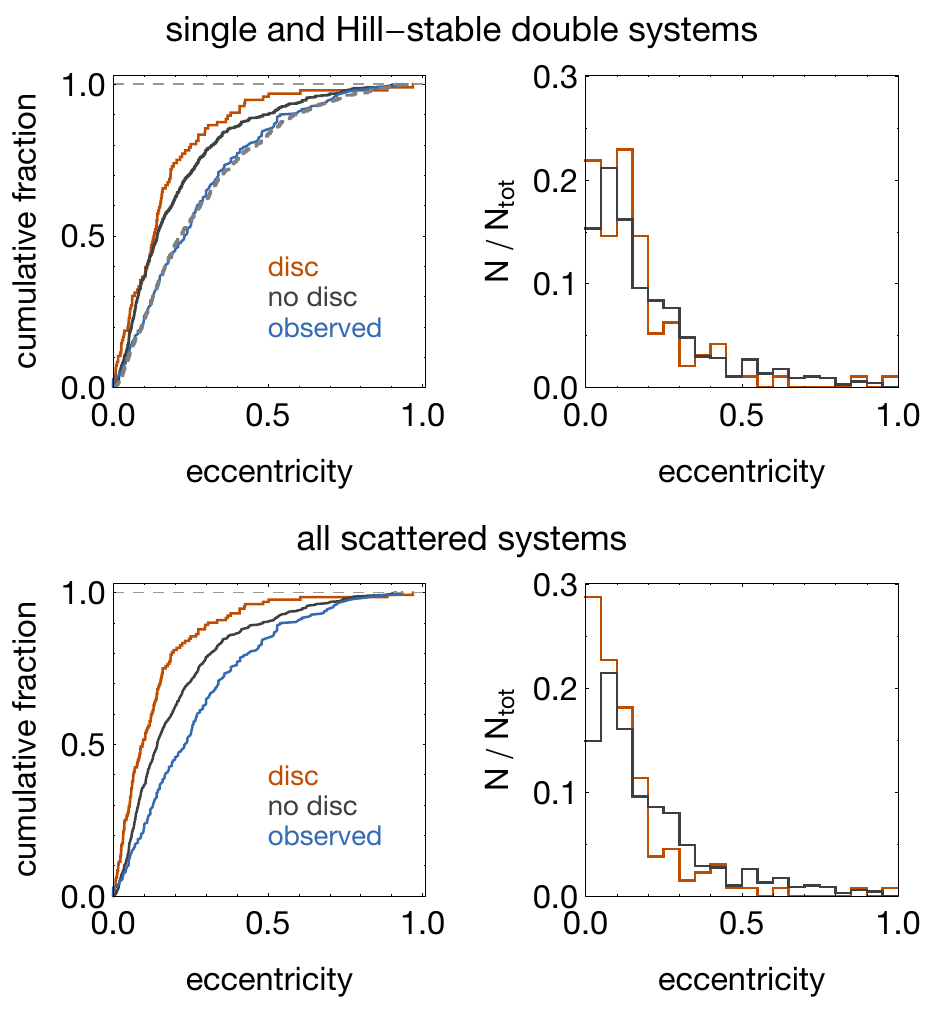}
 \caption{The eccentricity distribution at $10^8$ yr for observed exoplanets (blue), the runs with no disc (black), and the runs with a disc (red). The dashed line shows the disc-less result with random inclinations up to $1^\circ$. Top panels show only single and Hill stable double systems: K--S and A--D tests do not conclusively rule out the null hypothesis that the latter two distributions are the same (p-values 0.10 and 0.037, respectively). Bottom panels include all systems that scattered prior to 50 Myr, i.e. including triples and non Hill stable doubles. K--S and A--D tests reject the null hypothesis for these samples (p-values $< 10^{-3}$)}
 \label{eccentricity}
\end{figure}

\begin{figure}
 \includegraphics[width=84mm]{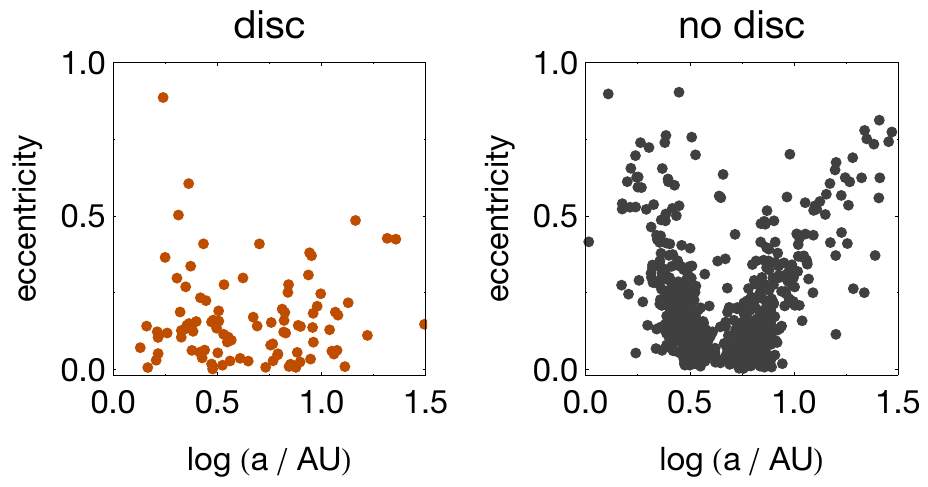}
 \caption{Semi-major axis--eccentricity scatter plots of the stable systems at $10^8$ yr with and without the disc. A two-dimensional K--S test rejects the null hypothesis that the two distributions are the same (p-value 0.041).}
 \label{aescatter}
\end{figure}

\begin{figure}
 \includegraphics[width=84mm]{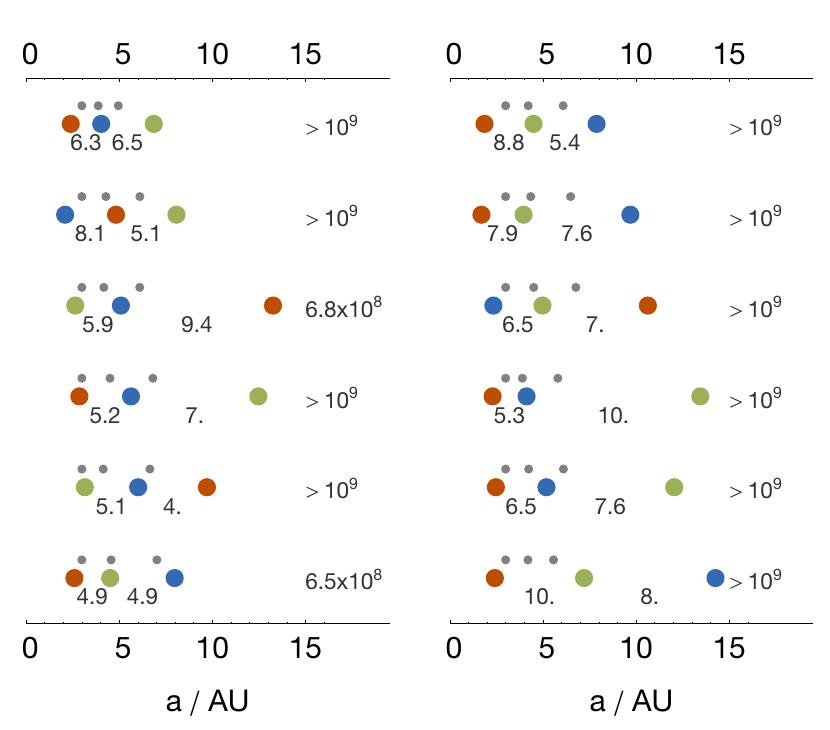}
 \caption{The spacings of the triple systems in the disc runs (the `$n_p=3$' bin in Figure \ref{instability_times}). Each row shows one system: small grey points show the initial spacing of the planets, and the large coloured points show the semi-major axes at 50 Myr. The colours correspond to the initial ordering: red points were originally the innermost planet, green points were in the middle, and blue points were outermost. The numbers between adjacent points show how many symmetric Hill radii separate the pair. The number to the right of each set shows the scattering time taken from extended runs to $10^9$ yr; only two systems went unstable in that time.
 }
 \label{stabletriples}
\end{figure}

\begin{figure}
 \includegraphics[width=84mm]{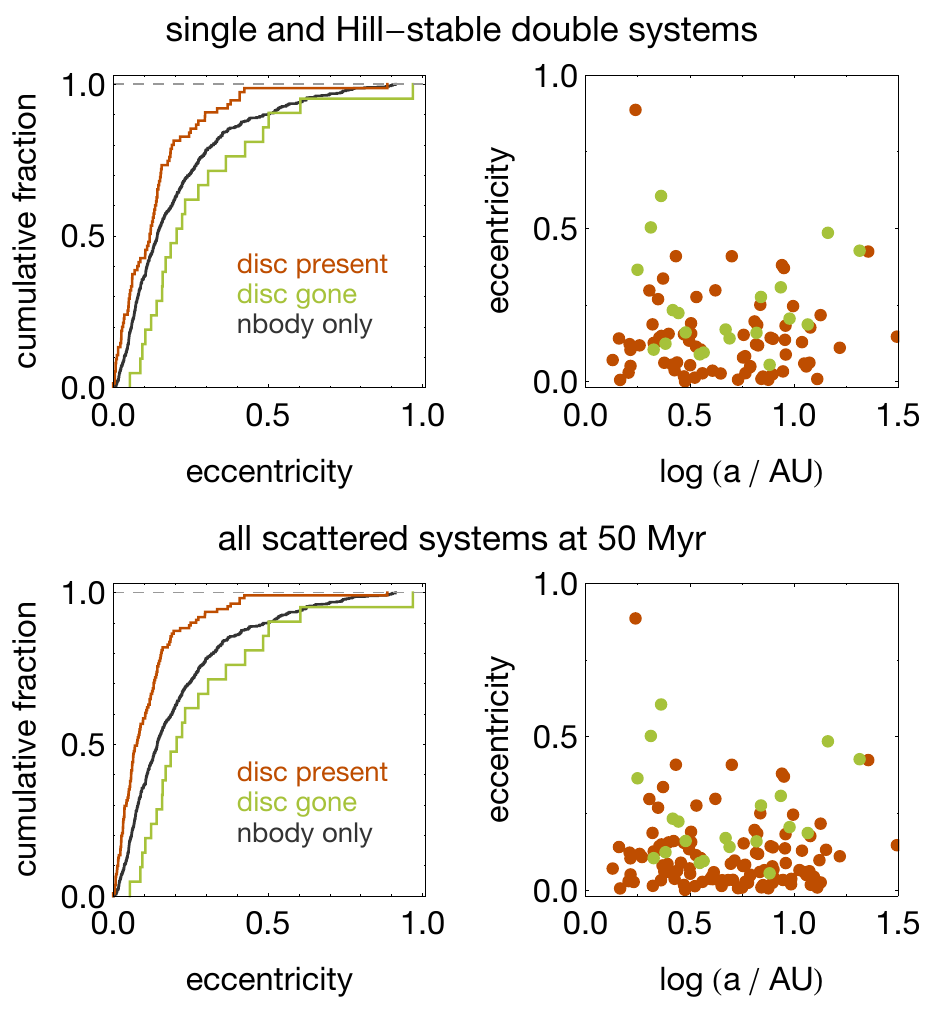}
 \caption{The eccentricity distribution of stable systems (top) and all scattered systems (bottom), with the runs with discs divided into those that scattered with disc material present and those that scattered after disc dispersal. K--S and A--D tests reject the null hypotheses that the early scattering systems follow the same distribution as the late scatterers (p-values 0.0016 and $3.2\times10^{-4}$) or the disc-less runs (p-values 0.0069 and $6.6\times10^{-4}$). The late scatterers are ambiguously distinct from the disc-less runs (p-values 0.067 and 0.033). Two-dimensional K--S tests on the semi-major axis--eccentricity distributions of the early and late scatterers likewise reject the null hypothesis only when unstable systems are included (p-value 0.071 with only stable systems, 0.0037 with unstable systems).}
 \label{early_late_comparison}
\end{figure}

With realistic planet masses, scattering from nearly (but {\em not} identically) coplanar and circular initial conditions is quite successful at reproducing the observed eccentricity distribution of giant extrasolar planets. Good agreement can be obtained irrespective of the initial number of planets \citep{ford08,chatterjee08,juric08}. These authors compare their results to the observed sample\footnote{Available from http://exoplanet.eu.}. In our study, a subtlety arises from the fact that  
\fargo is a 2D code, and thus we simulate the planetary dynamics strictly in the plane. This has some consequences for the eccentricity distribution obtained by purely gravitational studies; namely, the distribution is shifted toward lower eccentricities, clearly seen in Figure \ref{eccentricity}. We address this point further in appendix \ref{inclinationstudy}. When comparing our \fargo runs to a disc-less \nbody result we thus compare to a coplanar suite of experiments run with \mercury rather than directly to the observed sample. With non-zero inclinations, these initial conditions yield a distribution quite close to the observed exoplanets in a pure \nbody suite of experiments, which we verify by running the same initial conditions with random inclinations up to 1 degree. For the observed sample we exclude expolanets with $a<0.1$ au to take tidal circularisation out of consideration.

To statistically compare the sets of simulations we perform two-sample Kolmogorov--Smirnov (K--S) tests and two-sample Anderson--Darling (A--D) tests on the cumulative distributions. The latter test is more sensitive to deviations at the tails of the distribution. We reject the null hypothesis that the two samples are drawn from the same distribution when p-values are less than 0.05. In Figure \ref{eccentricity} we show the eccentricity distributions of the disc and disc-less cases. The top panels show the results including only those systems that are provably stable, i.e. single planet and Hill-stable double planet systems. The null hypothesis is not ruled out by the K--S (p-value = 0.10), though the A--D test rejects it (p-value = 0.037). The greater weight the A--D test gives to the distributions' tails validates the visual impression that the main difference is at the high eccentricity end. For comparison we also plot the observed eccentricity distribution, as well as the result of our simulations with inclinations up to 1 degree. While the eccentricity distributions alone are quite similar for stable systems, the combined eccentricity--semi-major axis distribution is marginally different. Figure \ref{aescatter} shows scatter plots of these distributions. A 2D K--S test \citep[using the method in][]{fasano87} rejects the null hypothesis, with a p-value of 0.041.

\begin{figure*}
 \includegraphics[width=180mm]{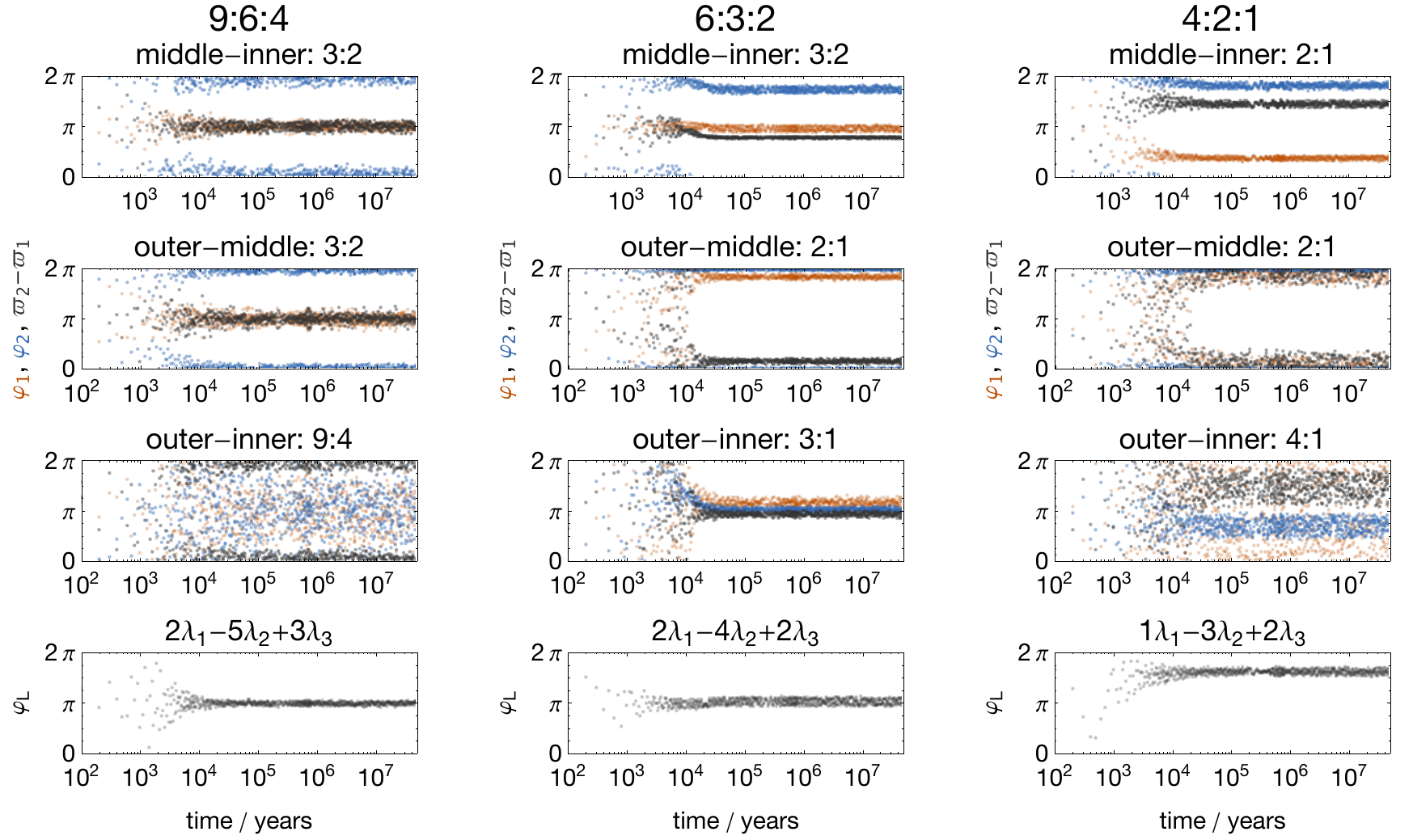}
 \caption{The evolution of the planetary system and the gas disc for an example run. The orbits of the planets are plotted on top of the surface density plots in the same colours as the plots of the orbital elements and masses. The surface density plot on the left shows the system before scattering, when the red, green, and blue planets have partially cleared a mutual gap. During the scattering event the green and blue planets collide to form the yellow planet on an orbit with $e \sim 0.6$. The aftermath of the scattering is shown in the middle density plot.  Over the next $\sim 2.5\times 10^4$ years, disc interactions damp the eccentricity of the outer planet; after disc clearing, the two planets are stable. The right density plot shows the system near the end of the photoevaporative clearing, with each planet orbiting in a gap. }
  \label{resonantchains}
\end{figure*}

The bottom panel of Figure \ref{eccentricity} shows the same comparison between the hydrodynamic and \nbody integrations, but now including all systems that scattered prior to 50 Myr, regardless of provable stability. This means that for the \fargo runs we include those triples that scattered and re-circularised in more widely spaced configurations (such as the one in Figure \ref{example2}), while for the \mercury runs we include the few double systems that are not provably stable. These distributions are visually more distinct, and this is born out by the statistical tests, with p-values for both tests less than $10^{-3}$. To test that the scattered triples are stable on longer timescales, we integrated the 12 cases further\footnote{For these integrations we used the Hybrid integrator of \mercuryns, with a 20 day stepsize.}, to $10^9$ yr. The spacing of these systems at 50 Myr, and the outcome of the further integration, are shown in Figure \ref{stabletriples}. With two exceptions, the systems remain stable for $10^9$ yr. 

Since the disc is the driver of any difference that might arise between the distributions, we split the sample into those in which instability begins when disc material is still present, and later scatterers. These distributions are shown in Figure \ref{early_late_comparison}. Whether or not we include the unstable systems, the distribution of early and late scatterers are significantly different, with K--S and A--D p-values of 0.0016 and $3.2\times10^{-4}$. While the early scatterers are likewise convincingly statistically distinct from the disc-less runs (p-values 0.0069 and $6.6\times10^{-4}$), the two tests also ambiguously reject the null hypothesis for the disc-less runs and the late scatterers (p-values 0.067 and 0.033). We would not necessarily expect any difference here, as there is no disc to alter the dynamics, and \citet{chatterjee08} showed that the time of instability is unimportant in determining the final outcome of a scattering experiment. Note that we suffer somewhat from low numbers of runs in this late scattering category, due to the large number of runs that never scatter over the length of our integrations.

\subsection{Mean Motion Resonances: 2 planets}
\label{MMR2}
Convergent migration in a disc can lead to trapping into resonance \citep[e.g.][]{lee02,snellgrove01,kley04,moorhead05}, and is seen with high frequency in the previous work most relevant to this one, the coupled \nbody--1D disc models of \citet{matsumura10}.
We first searched for mean motion resonances (MMRs) among the 39 stable double systems by observing the apsidal alignment $\varpi_2 - \varpi_1$ and the resonance angles for a $(p+q):q$ MMR given by \citep[e.g.][]{murray99}
\begin{equation}
\varphi_{1,2} = (p+q) \lambda_2 - p \lambda_1 - q \varpi_{1,2}.
\label{eq:resonanceangles}
\end{equation}	
Here subscript 1 and 2 refer to the inner and outer planet, $\lambda$ is the mean longitude, and $\varpi$ the longitude of pericenter. We searched for resonances up to 8th order by monitoring these angles for libration around fixed values. 

We find just two double systems in 3:1 MMRs with libration semi-amplitudes of $\sim 150$ and $\sim 80^\circ$. The case with the larger libration amplitude did not scatter until the disc had been photoevaporated. 
The low frequency of 2 planet resonances, and the large libration angles, are symptomatic of random entry into resonance 
due to purely gravitational effects \citep{raymond08}, rather than systematic driving into resonance due to gas disc 
migration \citep{matsumura10}. We observe a dramatically lower resonant fraction than the $\sim 75$~percent 
seen by \citet{matsumura10}. 
 The difference is probably due to the short lifetime of the discs in this study compared to the Myr dispersal timescales assumed by \citet{matsumura10}, which allows less time for post-scattering migration to move planets into resonance.

As noted in section \ref{instabilitytimesection}, the disc runs have a higher fraction that remain stable for the entire simulation than the \nbody only runs.
Because our initial conditions naturally place the planets near to several MMRs, there is the possibility that two adjacent planets may get trapped into resonance, a situation that can enhance the stability of a three planet system \citep{fabrycky10}. We also search each adjacent pair of planets in the systems that never scattered for MMRs. Out of 27 such cases, we find three in which the inner two planets are in resonance (two 3:2 and one 2:1), and three cases in which the outer two planets are in a 2:1 resonance. These resonant systems are potentially part of the reason for the large number of unscattered systems. 

\subsection{Mean Motion Resonances: 3 planet resonant chains}
\label{MMR3}
Further investigating the unscattered runs, we discovered that 12 of the 27 unscattered runs are in three body resonance. In eight cases the inner--middle and middle--outer pair move into resonance with librating resonance angles (equation \ref{eq:resonanceangles}), and in three of these the inner--outer pair are likewise in a resonance. The chains we find are 9:6:4 (four cases), 6:3:2, 3:2:1 (two cases), and 4:2:1\footnote{The Laplace resonance, as seen in GJ 876 \citep{rivera10} and potentially HD 82943 \citep{beauge08}.}. In each case a Laplace angle of the form 
\begin{equation}
\varphi_L=m \lambda_1 - (m+n) \lambda_2 + n \lambda_3 
\label{eq:laplaceangle}
\end{equation}
is identified \citep{aksnes88}, which librates around a fixed value for small integer values of $m$ and $n$, with subscripts 1,2,3 referring to the inner, middle, and outer planets.

The other four of the 12 triple resonances are less deeply in resonance; all three exhibit libration of the $m=1$, $n=3$ Laplace angle but the adjacent planet pairs are not in the 2:1 resonances. 
In Figure \ref{resonantchains} we plot the resonant behavior of three example chains; a representative of the 9:6:4 cases (i.e. a double 3:2 resonance), the 6:3:2 case, and the Laplace resonance case, the former two of which exhibit asymmetrical resonance in some of the angles. In the 9:6:4 run each adjacent pair of planets is trapped into a 3:2 resonance at $\sim4000$ yr, the same time that the Laplace angle begins librating around $\varphi_L=\pi$. The inner--outer pair move into apsidal alignment, but the resonant angles continue to circulate. 

The 6:3:2 case moves into triple resonance at $\sim1000$ yr, roughly the same time as the inner--middle pair lock into a 3:2 resonance, but somewhat before the other two pairings fully enter resonance at $\sim10^4$ yr. The initial inner--middle resonance and the Laplace angle all appear to be settling toward resonant angles of 0 or $\pi$, but when middle--outer and inner--outer resonance begin, most of the angles move into asymmetrical resonances, including the Laplace angle which shifts to librate around a value (very) slightly more than $\pi$. 

The double 2:1 Laplace resonance shows libration of $\varphi_L$ around an angle completely incommensurate with $\pi$. In this case the inner--middle 2:1 pair immediately enters asymmetric resonance, along with the Laplace angle. The middle--outer 2:1 pair are in a symmetric resonance, and the inner--outer pair are flirting with asymmetric resonance, though $\phi_1$ continues to circulate. We note that this asymmetric behavior in the Laplace resonance is a preferred outcome of simulations of migration into double resonance \citep{beauge08}, although this particular case did not undergo very much migration. It would appear that the high incidence of triple resonances, as well as the perhaps enhanced stability due to a single adjacent pair migrating into resonance, are the underlying reason behind the large number of unscattered systems in our hydrodynamic simulations compared to the \nbody runs.

\subsection{Periods of observability}
\label{observability}
\begin{figure}
 \includegraphics[width=84mm]{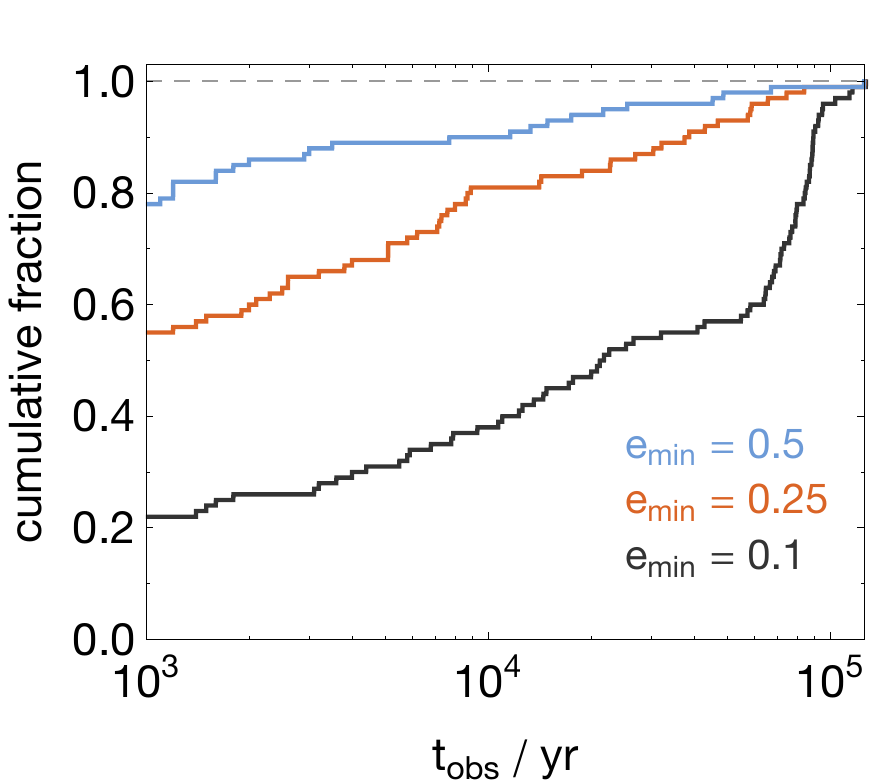}
 \caption{The cumulative distribution of $t_{obs}$, defined as the amount of time a system with at least 1 \mj~of gas left in the disc harbors at least one planet with $e > e_{min}$.} 
 \label{fig:tobs}
\end{figure}

Disc-planet interactions may be able to excite eccentricity to modest values, $e \sim 0.1$--0.2 \citep{dangelo06}, but planetary dynamics are probably necessary to reach higher values. Some of the more interesting features in the disc surface density in our simulations occur when eccentricities are high, during or just after scattering (see Figures \ref{example1} and \ref{example2}). These include non-axisymmetric features, new gaps being opened, and eccentric gaps. If scattering (or at least its signature in the eccentricity of the planets) is reasonably common in transitional discs, then these features may be observable. We estimate the likelihood that these features might be observed by calculating the amount of time that systems spend with at least one planet's eccentricity above a threshold value, while also having at least 1 \mj~ of gas left. This distribution is shown in Figure \ref{fig:tobs} for three values of the threshold eccentricity, $e_{min} = 0.1$, 0,25, and 0.5.

For the higher and more interesting thresholds, most of the runs do not have eccentric enough planets for any reasonable amount of time. If we require that interesting features are present for something like 25\% of our typical disc lifetime, $t_{obs} \sim 3 \times 10^4$ yr, a threshold of $e_{min}=0.25$ means that 10--15\% of the runs will have planets in eccentric enough orbits to be notieceable; thus a few percent of transition discs might be expected to harbor planets caught in the act or the aftermath of scattering.

\section{Conclusions}
\label{discussion}
If planets exit the gas disc phase in non-resonant configurations, a significant fraction of unstable systems are likely 
to scatter roughly contemporaneously with the transition between gas-rich and gas-poor conditions. We have studied 
whether the presence of small masses of gas, in photoevaporating transition discs, significantly modifies the outcome 
of scattering. Previous work that has addressed this question has been ambiguous. Detailed statistical studies, using 
approximate hydrodynamic treatments of the disc, have suggested that  gas free initial conditions are a reasonable 
approximation to the more physically complete case that includes gas \citep{chatterjee08,matsumura10}. On the 
other hand, earlier hydrodynamical work suggested a possibly significant effect \citep{moeckel08,marzari10}, but 
was hampered by very limited statistics. The immediate aftermath of a scattering event, when multiple
gravitationally interacting planets can have high-eccentricity
orbits that take them into previously undisturbed regions of
the disc, is clearly the most challenging situation to model
reliably with simple prescriptions. As shown, for example, in
Figure \ref{example1}, these phases can be important in shaping the long-term
evolution of scattered systems, though such individual examples
do not mean that the statistical results obtained from a
damping prescription might not be ``good enough".
Here, we have presented results from a direct hydrodynamic simulation 
of a moderately large ensemble of systems, which we have used to extract statistical results for the eccentricity 
distribution and for the abundance of resonant planetary configuration. We find some differences between
our hydrodynamic results and previously published results
derived using approximate methods, though we defer for future
work the question of whether {\em other} analytic prescriptions
for planet-disc interactions would yield better agreement.

Our main result is that scattering of random three-planet systems, from non-resonant initial conditions, yields 
significantly lower eccentricities (as compared to \nbody runs) when the effects of hydrodynamics are included. 
This difference does {\em not}, however, arise because of any major change in the dynamics of scattering events 
that either eject planets, or result in planetary mergers. In these --- dominant --- channels, scattering in the 
hydrodynamic and \nbody simulations leads to statistically very similar eccentricity distributions, differing 
(if at all) only in the fraction of highly eccentric planets. Rather, the lower eccentricities found in the 
hydrodynamic case occur because a modest number of scattering events evolve into three planet systems 
that appear to be stable over an observationally interesting time scale (at least a Gyr). These systems, which 
tend to have lower than average eccentricities, are simply not present in runs without gas, whose end state 
after scattering is uniformly either one or two planet systems.

Despite the lower eccentricities, we interpret our results as supporting the model of planetary scattering 
as the dominant physical mechanism giving rise to observed exoplanet eccentricity. Although gas does 
impact the outcome of scattering, it does not lead to large-scale circularization of planetary orbits. The 
differences between the hydrodynamic and gas-free simulations, in fact, are comparable to the 
variations seen in scattering simulations that start with different planetary mass functions \citep{raymond10}. 
It would almost certainly be possible to find some set of plausible initial conditions that would evolve 
hydrodynamically, under scattering, to match the observed distribution of massive planet eccentricity, 
given the freedom to fine tune the initial planetary spacing or disc dispersal time. Alternatively, 
and perhaps more naturally, the fraction of stable triple planet systems that are responsible for the 
lower eccentricity could likely be suppressed by starting with $n_p > 3$. Given the wealth of 
such possibilities, our results do not constitute an argument against the physical importance of 
scattering, though they do imply that the relatively precise agreement between \nbody 
calculations and observations is likely fortuitous.

The frequency of resonant systems of multiple planets is potentially a much more sensitive 
probe than the eccentricity distribution of the physics predating scattering. Under purely 
gravitational evolution, the occupation fraction of resonant systems mirrors the available 
phase space, and MMRs are uncommon \citep[a few percent,][]{raymond08}. In a laminar, 
viscous disc, conversely, almost arbitrarily high resonant fractions can be achieved 
\citep{matsumura10}. Our results fall somewhere in between these extremes. 
The frequency of 2-planet resonances that we find is comparable to that found in pure 
\nbody simulations. This is consistent with the conclusion, noted above, that the 
hydrodynamics and dynamics of scattering are very similar for those channels that 
result in planetary ejection or merger. However, we also observe a high frequency 
of resonant chains of 3-planet systems. Out of 100 runs, 12 evolved into libration of 
the Laplace angle $\varphi_L$. This resonance trapping stabilized the systems against instability, 
pushing the number of unscattered systems in our runs to nearly 30\%. Similar effects were 
seen in the \nbody simulations of \citet{raymond10}, where the dissipative role that is 
here played by gas was instead furnished by planetesimals. Observationally, some 
resonant systems are seen, but it is reasonably clear that their number is lower 
than theoretical expectations based on an extended period of evolution in a 
viscous disc. One possibility is that survival in resonance is frustrated at 
early times, when the disc is massive and turbulent fluctuations lead to stochastic 
torques on planets \citep{adams08}, such that resonance capture only becomes possible 
in a limited window just prior to disc dispersal. The quantitative details of such a 
scenario are, however, poorly known, and it is likely that our limited 
understanding of the mechanisms that can {\em disrupt} resonances is the dominant 
uncertainty in current models of early planetary system evolution.

Finally we have explored the potential of future imaging to directly observe the disc structures 
formed during the epoch of planetary scattering, which include eccentric gaps and massive 
planets directly accreting from the disc gas. The predicted structures are distinctive, 
but the limited synchronization between the onset of scattering and the dispersal 
of the disc means that they should only rarely be present in a sample of transition 
discs. We estimate an abundance that might be of the order of a few percent. A 
blind search of transition discs is thus unlikely to uncover any systems with 
ongoing scattering, though in principle it may be possible to exploit the large-scale 
asymmetries generated during scattering to identify possible candidates.

\section*{Acknowledgments}
Our thanks to the referee for a quick and constructive report.
This work was performed in part using the Darwin Supercomputer of the University of Cambridge High Performance Computing
Service (http://www.hpc.cam.ac.uk/), provided by Dell Inc. using Strategic Research Infrastructure Funding from the Higher
Education Funding Council for England. PJA acknowledges support from NASA, under award NNX09AB90G and NNX11AE12G from the Origins of Solar Systems and 
Astrophysics Theory programs, and from the NSF under award AST-0807471.

\bibliographystyle{mn2e}

\appendix
\section{The effect of zero-inclination initial conditions}
\label{inclinationstudy}
The \fargo and \mercury scattering calculations reported in this paper employ exactly 
planar planetary initial conditions. This is done for reasons of computational economy: 
3-dimensional hydrodynamic simulations with the same in-plane resolution as the 2D runs 
would require around two orders of magnitude greater resources. For many purposes, moreover, 
assuming planarity is not a bad approximation. Prior $n$-body scattering calculations, run 
using the same mass distribution as ours, show that the median inclination of surviving 
planets is only $\approx 10^\circ$ \citep{raymond10}. 
This means that the initial phase of scattering 
typically involves inclinations $i \sim h/r$, or smaller, and accordingly we expect that 
any hydrodynamic interaction with the disc will be dominated by planar effects (eccentricity 
damping, accretion) rather than fully 3D ones (such as warp excitation).
As noted in Section \ref{results}, however, the assumption of exact coplanarity does 
introduce a technical subtlety, since the final $f(e)$ that results from scattering in 
planar systems is not the same as that which results from otherwise identical initial 
conditions with a small (of the order of one degree) non-zero inclination. Because of 
this subtlety, we compare our hydrodynamic simulations not to the same $n$-body runs 
that match the observed eccentricity distribution, but rather to their zero inclination 
analogues. 

We emphasize that the point of this paper is not to argue that the result of the planar 
calculation is physically realistic. Rather, from the fact that in 2D the effect of 
hydrodynamics on the final eccentricity distribution is small, we infer that in 3D 
a hydrodynamic calculation would yield results similar to the observed distribution. 
That said, a physical effect is in principle possible if the gas disc were able to damp 
mutual inclinations\citep{lubow01,marzari09} to the point 
where the final system was effectively planar. To quantify this possibility, we have 
run a set of $n$-body \mercury experiments to identify how small the initial inclination 
must be for the system to scatter as if it were exactly planar. The setup is identical 
to that used earlier. We start with three planets on circular orbits, separated by 
$K = 4.0$ mutual Hill radii (Equation~\ref{eq:Hill}), with the inner planet at 3~AU.  
The mass spectrum is $f(m) \propto 
m^{-1.1}$ between $0.3 \ M_J$ and $5 \ M_J$, and the planetary radius is $1.3 \ R_J$. 
We run four ensembles of 500 simulations, with the planetary inclinations chosen randomly and 
uniformly in the range $i = [0,1^\circ]$, $i = [0,0.1^\circ]$, $i = [0,0.01^\circ]$ and 
$i=0$. The systems are integrated for 15~Myr, after which we plot the eccentricity 
distribution of the surviving planets in the unstable systems (at this epoch, this is 
about 80\% of the total). As before, we discard double systems that are not Hill stable, which removes only $\sim 10$ runs from each set.

\begin{figure}
 \includegraphics[width=84mm]{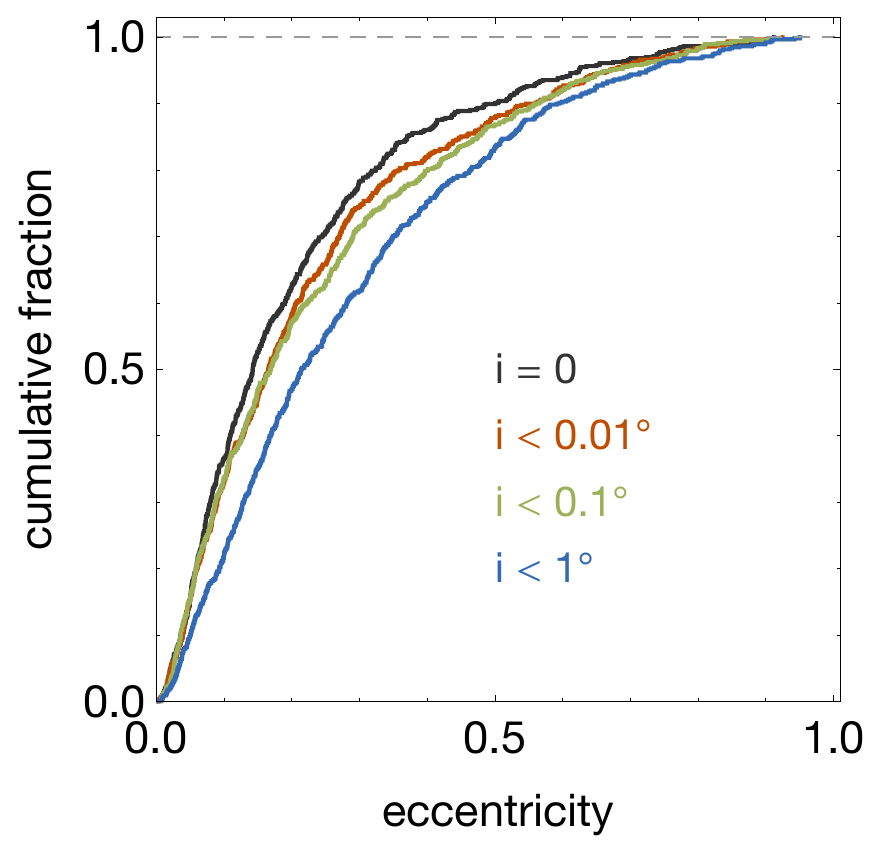}
 \caption{The cumulative eccentricity distributions, following scattering, obtained from purely $n$-body simulations of
 3-planet systems initialized with different inclinations. The highest mean eccentricities result from initial conditions in
 which all three planets had inclinations chosen randomly and uniformly in the interval $i = [0,1^\circ]$ (blue curve).
 Lower eccentricities result from choosing $i = [0,0.1^\circ]$ (green), or $i = [0,0.01^\circ]$ (red). The black curve
 shows results for a strictly planar geometry. In all cases, only surviving planets from those systems that went unstable 
 during the integration are plotted.} 
 \label{fig:incstudy}
\end{figure}

The results for the eccentricity distribution are shown in Figure~\ref{fig:incstudy}. We find 
that the ensembles initialized with $i = [0,0.1^\circ]$ and $i = [0,0.01^\circ]$ are similar to, 
but statistically distinct from, the zero inclination limit (K--S p-values 0.049 and 0.023), whereas inclinations of the order of 1~degree 
are sufficient to result in markedly larger final eccentricities, distinct from all other distributions. To an order of magnitude, 
the condition for scattering behavior that is distinct from the 3D limit in these three planet systems is thus that 
the initial inclinations should satisfy $i < 0.1^\circ$.

Could real planetary systems be flat enough that their scattering dynamics are effectively 
two-dimensional? It seems unlikely. Although a gas disc damps the inclination of a single planet 
on a short time scale, it will damp the orbit until it is coincident with the local angular momentum 
vector of the gas, which will not be exactly constant with radius. A warp whose magnitude (over several AU)   
exceeds the threshold $\Delta i = 0.1^\circ$ seems almost inevitable \citep{terquem00,lubow10}. Even if the gas disc were perfectly flat, any temporary resonances between 
the planets would likely excite their mutual inclination \citep{thommes03} 
to the level where subsequent scattering would occur in the three-dimensional limit.

\bsp

\label{lastpage}

\end{document}